\documentstyle[aps,prb,preprint,tighten]{revtex}
\begin{document}
\draft

\title{Stability, dynamical properties and melting of a
classical bi-layer Wigner Crystal}

\author{G. Goldoni\cite{byline1} and F. M. Peeters\cite{byline2}}

\address{Departement Natuurkunde, Universiteit Antwerpen (UIA), 
Universiteitsplein 1, B-2610 Antwerpen, Belgium}

\date{\today}
\maketitle

\begin{abstract}

 We investigate the stability, the dynamical properties and
melting of a two-dimensional (2D) Wigner crystal (WC) of classical
Coulombic particles in a bi-layer structure. Compared to the
single-layer WC, this system shows a rich phase diagram.  Five
different crystalline phases are stable; the energetically
favoured structure can be tuned by changing either the inter-layer
distance or the particle density. Phase boundaries consist of both
continuous and discontinuous transitions. We calculated the phonon
excitations of the system within the harmonic approximation and we
evaluated the melting temperature of the bi-layer WC by use of a
modified Lindemann criterion, appropriate to 2D systems. We
minimized the harmonic free-energy of the system with respect to
the lattice geometry at different values of
temperature/inter-layer distance and we found no
temperature-induced structural phase transition. 

\end{abstract}


\narrowtext

\section*{introduction}
\label{sec:introduction}

Classical charged particles confined in a single, two-dimensional
(2D) layer localize into a hexagonal lattice (Wigner crystal) for
sufficiently large densities and low
temperatures.~\cite{Isihara89} Such a single-layer Wigner crystal
(SLWC) has been realized, e.g., on the surface of liquid
helium.~\cite{Grimes79} Colloidal particles dissolved in water and
placed between two glass plates is another example of an
experimental system where classical particles exhibit Wigner
crystallization.~\cite{Murray87} Electrons in 2D semiconductor
heterostructures behave like quantum particles, but a strong
perpendicular magnetic field quenches the kinetic energy, leading
the system toward the classical regime.  The quest for the
observation of such a Wigner crystal has been the object of very
intense work over the last decades.~\cite{Clark91} The melting
transition of the classical SLWC, in particular, has attracted a
large body of investigation~\cite{Strandburg88} since the proposal
of a dislocation-mediated melting mechanism, leading to the
prediction of a continuos melting
transition.~\cite{Kosterlitz73,Halperin78,Young79}

Recently, a new 2D system has attracted the attention of several
groups, namely, the Wigner crystal in a bi-layer structure. One of
the peculiarities of the bi-layer Wigner crystal (BLWC), compared
to the SLWC, consists in the rich phase diagram; it has been
predicted that different crystalline structures are stable in
different ranges of inter-layer distance/charge density.
\cite{Vilk84,Vilk85,Falko94,Esfarjani95,Goldoni95,NarasimhanPre,ChanPre}

In the present paper we address the phase diagram of such bi-layer
structures.~\cite{Goldoni95} We consider a BLWC of Coulombic
particles evenly distributed between the two layers.  In a
classical BLWC, the inter-particle interaction can be
characterized by a unique dimensionless parameter
$\eta=d\sqrt{n/2}$, where $d$ is the inter-layer distance and $n$
is the total charge density.  $\eta$ represents the ratio between
the inter-particle interaction {\em between} the layer and {\em
within} each layer.  Thus, in the classical case, the Hamiltonian
of the system is only a function of $\eta$, which therefore
determines completely the phase diagram at $T=0$. This is in
contrast with the equivalent quantum problem, where $d$ and $n$ do
not scale out. The search for the stable structure of a classical
BLWC, at $T=0$ and as a function of $\eta$, is made easier by the
following considerations: 1) due to the long-range interaction,
the two lattices which occupy the two layers (sub-lattices) are
staggered to maximize the inter-particle distance. Each lattice
site sits at the center of a cell in the opposite layer; 2) there
are two trivial limiting cases: at $\eta=0$ the two sub-lattices
reduce to a SLWC, which is known to crystallize in a hexagonal
lattice (one-component hexagonal lattice). At the opposite limit
of large $\eta$ the two sub-lattices are weakly coupled and,
therefore, the stable structure is constituted by two staggered
SLWC (staggered hexagonal lattice).  By comparing the static
energy of several lattices, we find that five different phases are
energetically favoured in different ranges of $\eta$. The five
structures, in order of increasing $\eta$, are a one-component
hexagonal lattice (I), a staggered rectangular lattice (II), a
staggered square lattice (III), a staggered rhombic lattice (IV),
and a staggered hexagonal lattice (V).  These phases evolve one
into the other through both first and second order phase
transitions. 

There exist already a number of investigations of the $T=0$ phase
diagram of the classical BLWC in various
systems.~\cite{Vilk84,Falko94,Esfarjani95,Goldoni95,ChanPre} Some
of the previous investigations of the present
system~\cite{Falko94,ChanPre} did not identify all five phases. In
Ref.~\onlinecite{Esfarjani95}, the bi-layer electron system which
forms in a single wide quantum well above a critical
density~\cite{Suen92} was studied. In this case, the transition
from the single layer to the bi-layer (and, at higher densities,
to a higher number of layers) is of first order, with $\eta$ which
jumps from $0$ to $\sim0.27$; therefore, the low-$\eta$ phases I
and II were not investigated in Ref.~\onlinecite{Esfarjani95}.
Theoretical investigations suggest that also bi-layer structures
in the quantum regime possess a complex phase diagram. In
Ref.~\onlinecite{Chen93} a structural instability is found in the
strong coupling regime, when the tunneling probability decreases
below a critical threshold as a consequence of layer separation.
The five phases described above have been predicted to exist in a
bi-layer quantum Hall system.~\cite{NarasimhanPre} In this case,
the phase diagram is even more complex, because a spin structure
is associated with the lattice structure.  Other bi-layer
structures with an even more complex phase diagram can be
imagined, for example, with different densities of the two
layers.~\cite{Vilk84,Vilk85}

 Phonon excitations of some of the above systems have been
partially investigated, within the harmonic
approximation.~\cite{Falko94,Esfarjani95,ChanPre} Phonon
frequencies have been used to evaluate the zero-point energy and
the critical density for cold melting of the quantum
BLWC~\cite{Esfarjani95,ChanPre} also in the presence of a magnetic
field. 

In this paper we address the {\em non-zero} temperature phase
diagram of a classical BLWC, which has not been investigated so
far. With this aim, we have systematically investigated, within
the harmonic approximation, the phonon excitations of each of the
five structures. This allowed us not only to determine the range
of structural stability and the behaviour of the acoustical and
optical modes, which could be subject to experimental
observations, but also allowed us to estimate the melting
temperature of the crystal through the Lindemann criterion. The
latter states that melting takes place when the mean square
displacement of the crystallized charges exceeds a certain
fraction of the lattice parameter. It should be noted that in 2D
the mean square displacement diverges logarithmically with the
crystal size. On the other hand, the {\em relative} mean square
displacement between nearest neighbours (NN's) is a well defined
quantity and its value at melting has been determined from
simulations of 2D crystals.~\cite{Bedanov85} Therefore, the latter
quantity has been used in the present paper to estimate the
melting temperature. Our results show that, by changing $\eta$ at
constant temperature, one can pass through alternating regions of
crystalline and liquid order. 

The overlapping range of stability of different lattice structures
along the $\eta$ axis suggests the possibility that
temperature-induced structural phase transitions take place,
before the melting temperature is reached. To investigate this
last possibility, we have minimized the harmonic free-energy of
different structures at increasing temperatures at fixed $\eta$,
and we have found no evidence of such a temperature-induced
structural phase transition. 

The paper is organized as follows. In Sec.~\ref{sec:static} we
investigate the zero temperature phase diagram. In
Sec.~\ref{sec:dynamic} we calculate the phonon excitations of the
systems. The non-zero temperature phase diagram is investigated in
Sec.~\ref{sec:finiteT}. Results are discussed and summarized in
the last section. 

\section{Zero temperature phase diagram}
\label{sec:static}

We consider a BLWC consisting of $N$ classical, spinless particles
with total charge density $n$, evenly distributed over the two
layers.  The same form of the Coulombic interaction $e^2/r$ is
assumed between particles in the same layer and in different
layers. 

Electrons crystallized in the two layers constitute two
sub-lattices which are equivalent by symmetry. When necessary, we
denote the two sub-lattices by A and B. We consider only the case
of two layers of equal charge density $n_s=n/2$ each;  therefore, in 
the limit $\eta\rightarrow 0$, we recover a
SLWC of density $n$.  We consider the BLWC as a 2D lattice of
$N/2$ unitary cells, with two electrons per cell  sitting on 
opposite layers.  The primitive vectors are denoted by 
${\bf a}_1$ and ${\bf a}_2$, and the basis vectors are $(0,0)$  and 
${\bf c}$; ${\bf a}_1$, 
${\bf a}_2$, ${\bf c}$, $n_s$, and the vectors ${\bf b}_1$ and 
${\bf b}_2$ generating the reciprocal lattice,  are listed in 
Table \ref{geometries} for the five relevant phases.  The 
equilibrium positions of the 
crystallized electrons in layer A and B are, respectively, 
\begin{eqnarray} 
{\bf R}_A & = & i {\bf a}_1 + j {\bf a}_2 \\ 
{\bf R}_B & = & i {\bf a}_1 + j {\bf a}_2 + {\bf c} 
\end{eqnarray} 
where $i,j$ are integers. The total potential energy of the mobile 
charges due to the intra-layer and the inter-layer 
interaction is
\begin{eqnarray}
V_p &=& \frac{1}{2} 
   \left\{
   \sum_{{\bf R}_A\neq{\bf R}^\prime_A}\frac{e^2}
{\left|{\bf R}_A-{\bf R}^\prime_A\right|} 
  +\sum_{{\bf R}_B\neq{\bf R}^\prime_B}\frac{e^2}
{\left|{\bf R}_B-{\bf R}^\prime_B\right|}
 +2\sum_{{\bf R}_A,{\bf R}_B}     \frac{e^2}{
           \left[\left|{\bf R}_A-{\bf R}_B\right|^2+d^2\right]^{1/2}} 
   \right\}.
\label{potential_energy} 
\end{eqnarray}
The factor $1/2$ accounts for double counting.
Since the two layers are equivalent, and the origin can be chosen 
arbitrarily if we neglect surface effects, there are $N/2$ 
equivalent terms in each sum and  the potential energy per 
particle $E=V_p/N$, therefore,  reads 
\begin{equation}
E = \frac{1}{2}\left(E_0+E_I\right) , 
\end{equation}
where
\begin{equation}
E_0 = \sum_{{\bf R}\neq0}\frac{e^2}{R}
\label{E0}
\end{equation}
represents the intra-layer interaction energy, and
\begin{equation}
E_I = \sum_{{\bf R}}\frac{e^2}
  {\left[\left|{\bf R}+{\bf c}\right|^2+d^2\right]^{1/2}}
\label{Ei}
\end{equation}
is  the inter-layer interaction energy. Here ${\bf R}=i{\bf 
a}_1+j{\bf a}_2$  and $R=\left|{\bf R}\right|$.

We follow Bonsall and Maradudin~\cite{Bonsall77} and rewrite 
(\ref{E0}) as
\begin{equation}
E_0 = e^2 
\lim_{r\rightarrow0}\left[\sum_{\bf R}\frac{1}{\left|{\bf r}-
{\bf R}\right|}
-\frac{1}{r}\right] ,
\label{E0BM}
\end{equation}
where ${\bf r}$ is a 2D vector and $r=|{\bf r}|$.
It is convenient to define the following two functions~\cite{Fisher82}
\begin{eqnarray}
T_0\left({\bf r},{\bf q}\right)&=&
e^{-i{\bf q}\cdot{\bf r}}\sum_{\bf R} \frac{
e^{i{\bf q}\cdot\left({\bf r}-{\bf R}\right)}}
{\left|{\bf r}-{\bf R}\right|}
-\frac{1}{r} , 
\label{def_t0}\\
T_I\left({\bf r},{\bf q}\right)&=&
e^{-i{\bf q}\cdot{\bf r}}\sum_{\bf R} \frac{
e^{i{\bf q}\cdot\left({\bf r}-{\bf R}+{\bf c}\right)}}
{\left[\left|{\bf r}-{\bf R}+{\bf c}\right|^2+d^2\right]^{1/2}} ,
\label{def_ti}
\end{eqnarray}
from which $E_0$ and $E_I$ are obtained
\begin{eqnarray}
E_0 & = & e^2 \lim_{r\rightarrow0} T_0({\bf r},0) ,
\label{E_0} \\
E_I & = & e^2 T_I(0,0) .
\label{E_I} 
\end{eqnarray}
Due to the long range nature of the interaction, the lattice sums 
in $T_0$ and $T_I$ converge slowly. The Ewald technique 
is commonly used to overcome this difficulty, and consists in 
splitting the slowly convergent sum into two parts: the 
contribution of the first shells of neighbours is summed up in 
real space, while the contribution of the outer shells is 
summed up in reciprocal space. Both sums turn out to be 
rapidly convergent. The transformation 
to rapidly convergent sums over the real 
lattice vectors ${\bf R}$ and the reciprocal lattice vectors ${\bf G}$  
is reported in Appendix \ref{sec:appA}. 
Here we state the final result
\widetext
\begin{eqnarray}
T_0({\bf r},{\bf q})&=&
\sqrt{n_s}\sum_{{\bf G}} e^{-i\left({\bf q}+{\bf G}\right)\cdot{\bf r}}
      \Phi\left(\frac{\left|{\bf q}+{\bf G}\right|^2}{4\pi n_s}\right)
+\sqrt{n_s}\sum_{{\bf R}\neq0} e^{-i{\bf q}\cdot{\bf R}}
          \Phi\left(\pi n_s\left|{\bf r}-{\bf R}\right|^2\right)
\nonumber\\
&& +\sqrt{n_s}\Phi\left(\pi n_s r^2\right)-\frac{1}{r} , 
\label{T0}\\
T_I({\bf r},{\bf q})&=&
\sqrt{n_s}\sum_{{\bf G}} e^{-i\left({\bf q}+{\bf G}\right)\cdot{\bf r}}
          e^{-i{\bf G}\cdot{\bf c}}
          \Psi\left(\frac{\left|{\bf q}+{\bf G}\right|^2}{4\pi n_s},
                    \pi\eta^2\right) \nonumber\\
&&+\sqrt{n_s}\sum_{{\bf R}}
e^{-i{\bf q}\cdot\left({\bf R}-{\bf c}\right)}\Phi\left(\pi 
\left[n_s\left|{\bf r}-{\bf R}+{\bf c}\right|^2+\eta^2\right]\right) .
\label{Ti}
\end{eqnarray}
\narrowtext\noindent
The functions $\Phi(x)$ and $\Psi(x,y)$, defined in Appendix
\ref{sec:appA}, decay exponentially to zero for large  
$x$; therefore, $T_0$ and $T_I$ contain only rapidly convergent 
sums.

The $G=0$ term in the first
sum on the rhs of (\ref{T0}) and (\ref{Ti}) give rise to a  
divergent term in $E_0$ and in $E_I$, ensuing from
the lack of charge neutrality. These 
terms are balanced by the interaction with a positive 
background, independently from the lattice geometry. Since the 
origin of the energy can be chosen arbitrarily, the divergent 
terms can be separated out and neglected in the calculation of 
the energy. This is done in Appendix~\ref{sec:AppB} and the final 
result is
\widetext
\begin{eqnarray}
E_0 &=& 2 e^2 \sqrt{n_s}\left\{\sum_{{\bf R}\neq0}\Phi\left(\pi 
n_sR^2\right)-2\right\}, \label{e0_explicit}\\
E_I &=&  e^2 \sqrt{n_s}\left\{\sum_{\bf R}
\Phi\left(\pi\left[n_s\left|{\bf R}+{\bf 
c}\right|^2+\eta^2\right]\right)
\right.\nonumber\\
&&\left.+\sum_{{\bf G}\neq0}e^{i{\bf G}\cdot{\bf c}}
\Psi\left(\frac{G^2}{4\pi n_s},\pi\eta^2\right)
+2\left\{\pi\eta\,\mbox{erfc}\sqrt{\pi}
\eta-e^{-\pi\eta^2}\right\}\right\},
\label{Ei_explicit}
\end{eqnarray}
\narrowtext\noindent
where  $G=|{\bf G}|$ and $\,\mbox{erfc}(x)$ is the complementary 
error function defined in Appendix~\ref{sec:appA}.

We have calculated $E/e^2\sqrt{n}$ (recall that $n=2n_s$) for the
five different lattices listed in Table \ref{geometries}. Phases
I, III, and V are ``rigid'', meaning that, for a fixed density,
the cell is uniquely determined. On the contrary, phases II and IV
are ``soft'', because each of them contains a parameter, the ratio
$a_2/a_1$ of the length of the primitive vectors (aspect ratio)
and the angle between them, respectively, which can take on
continuous values; at each value of $\eta$, this parameter is
determined by energy minimization. 

Figure \ref{fig:totene} shows the calculated energy per particle
of phases I, III, and V as a function of $\eta$; the energy of
phases II and IV would be nearly indistinguishable on the scale of
this figure and will be shown later in Fig.~\ref{fig:totene2}. As
we discussed in the introduction, phase I is energetically
favoured at very small $\eta$, and it reduces to the SLWC at
$\eta=0$. At $\eta=0$ we find $E=-1.96052\, e^2\sqrt{n}$, which
coincides with Ref.~\onlinecite{Bonsall77}. At the opposite limit
of large $\eta$ the two sub-lattices become less and less coupled
and the favoured geometry is composed of two staggered SLWC (phase
V). Accordingly, the energy converges to the value
$E_0=-1.96052\,e^2\sqrt{n}/\sqrt{2}$, where the factor $\sqrt{2}$
accounts for the reduced charge density. In the intermediate range
of $\eta$ the energetically favoured structure is phase III. The
geometry of the three phases is sketched in the insets of
Fig.~\ref{fig:totene}, close to the range where they have the
lowest energy. 

Within the regions delimited by the open dots in
Fig.~\ref{fig:totene}, the energetically favoured structures are
phases II and IV. These intermediate phases allow the lattice to
pass from phase I to phase III, and from phase III to phase V,
respectively. Note that phase II contains phase I and phase III as
limiting cases, corresponding to the aspect ratios
$a_2/a_1=\sqrt{3}$ and $a_2/a_1=1$, respectively. Analogously,
phase IV contains phase III as limiting case for $\theta=\pi/2$.
Therefore, the transitions I$\rightarrow$II, II$\rightarrow$III,
and III$\rightarrow$IV are continuous. Note, on the other hand,
that there is no continuous way to pass from phase IV to phase V
and, therefore, the transition IV$\rightarrow$V is of the first
order, i.e., $\partial E/\partial\eta$ exhibits a jump. 
Apparently, this point has been overlooked in
Ref.~\onlinecite{Esfarjani95}, where the authors claim that the
transition IV$\rightarrow$V is continuous. The necessity for a
first order phase transition when going from phase III to phase V
has also been discussed, by use of general group theoretical
arguments, by V. Fal'ko in Ref.~\onlinecite{Falko94}. 

Figure~\ref{fig:totene2}(a) shows the transition
I$\rightarrow$II$\rightarrow$III on an enlarged scale. Phase I is
energetically favoured only in a very small range around $\eta=0$.
As $\eta>0.006$, in fact, a rectangular unit cell with
$a_2/a_1<\sqrt{3}$ (phase II) is energetically favoured. In the
inset of Fig.~\ref{fig:totene2}(a) we show how the aspect ratio
$a_2/a_1$ evolves in a continuous way during the transition; 
phase I evolves into phase III through an anisotropic shrinking of
the rectangular unit cell, and eventually $a_2/a_1=1$,
corresponding to phase III, is reached at $\eta=0.262$. 

The energy of phase IV is compared in Fig.~\ref{fig:totene2}(b)
with the energy of phase III and V.  For $0.622<\eta<0.732$ the
staggered rhombic lattice (phase IV) has the lowest energy. As
shown in the inset, increasing $\eta$ the angle $\theta$ between
the cell axes evolves continuously from $90^\circ$, corresponding
to the square lattice (phase III), to $69.48^\circ$, and suddenly
drops to $60^\circ$, which corresponds to phase V.  The phase
boundaries found above agree well with those found in
Refs.~\onlinecite{Falko94} and \onlinecite{Esfarjani95}. 

To conclude this section, we give the asymptotic expressions of 
the static energy, for small and large $\eta$,
\widetext
\begin{eqnarray}
\frac{E}{e^2\sqrt{n}}  & = &  -1.96052 + \frac{1}{2\sqrt{2}}
\left[2\pi\eta-0.600434\,\eta^2+2.86713\,\eta^4\right] 
\hspace{0.5truecm} \mbox{for small}\, \eta \label{small_eta},\\
\frac{E}{e^2\sqrt{n}}  & = &  -\frac{1.96052}{\sqrt{2}} -
\frac{v}{2\sqrt{2}} e^{-w\,\eta}
\hspace{0.5truecm} \mbox{for large}\, \eta \label{large_eta},
\end{eqnarray}
\narrowtext\noindent
where $v=-3(\sqrt{3}/2)^{1/2}$ and $w=(8\pi^2/\sqrt{3})^{1/2}$ in 
(\ref{large_eta}). Eqs.~(\ref{small_eta}) and (\ref{large_eta}) 
reproduce the correct energy  within $2\,\%$ for $\eta<0.3$ and 
within 
$0.2\,\%$ for $\eta>0.5$, respectively. We stress that the above 
expressions are not 
fitting functions, but have been obtained by a series expansion of 
(\ref{Ei_explicit}) with respect to $\eta$ in the relevant range. 
In Eq.~(\ref{small_eta}) the linear term, ensuing from the last 
term in Eq.~(\ref{Ei_explicit}), is the only 
odd order term in the Taylor 
expansion and all higher order terms are even; the coefficients 
involve 
sums over the direct and the reciprocal lattice, which have 
been calculated numerically for the lattice of phase I. 
The coefficients in Eq.~(\ref{large_eta}) can be obtained 
analytically, once one realizes that, for large $\eta$, only
the first shell of ${\bf G}$'s needs to be retained in 
(\ref{Ei_explicit}); higher order terms are proportional to 
$e^{-\pi\eta^2}$ and decay faster for large $\eta$. In 
Ref.~\onlinecite{Esfarjani95} two fitting expressions for the 
classical energy were given; however, we found that none of them 
have the correct limiting behaviour. 

\section{dynamical properties}
\label{sec:dynamic}

In this section we calculate the frequencies of the phonon
excitations of the five different phases within the harmonic
approximation. 

For a general lattice, the square of the phonon frequencies are the 
eigenvalues of the dynamical matrix defined by~\cite{MaradudinSSP}
\begin{eqnarray}
\left[{\bf 
T}({\bf q};l\kappa, l^\prime\kappa^\prime)\right]_{\alpha\beta} &&=
\frac{1}{\left(m_\kappa m_{\kappa^\prime}\right)^{1/2}} \nonumber\\
&&\sum_{l^\prime}\phi_{\alpha\beta}\left(l\kappa, 
l^\prime\kappa^\prime\right)
e^{-i{\bf q}\cdot\left({\bf R}_{l\kappa}-{\bf 
R}_{l^\prime\kappa^\prime}\right)}, \label{gen_dyn_mat}
\end{eqnarray}
where ${\bf R}_{l\kappa}$ is the position 
vector of the $\kappa$-th particle in the $l$-th cell of the crystal,
and $m_\kappa$ its mass. The quantities
$\phi\left(l\kappa, l^\prime\kappa^\prime\right)$  are the force 
constants defined by
\begin{equation}
\phi_{\alpha\beta}\left(l\kappa, l^\prime\kappa^\prime\right)=
\partial_\alpha\partial_\beta
\phi\left({\bf R}_{l\kappa}-{\bf R}_{l^\prime\kappa^\prime}\right) ,
\end{equation}
where 
$\phi\left({\bf R}_{l\kappa}-{\bf R}_{l^\prime\kappa^\prime}\right)$ 
is the
two-body inter-particle potential. Here and in the following we
use the notation $\partial_\alpha F({\bf x})=\partial
F({\bf x}^\prime)/\partial x_\alpha^\prime|_{{\bf x}={\bf x}^\prime}$,
where 
$x_\alpha$ is the $\alpha$-th component of the vector ${\bf x}$.
Due to translational invariance, the force 
constants satisfy the sum rule~\cite{MaradudinSSP}
\begin{equation}
\sum_{l\kappa, l^\prime\kappa^\prime} 
\phi_{\alpha\beta}\left(l\kappa, l^\prime\kappa^\prime\right) =0.
\label{sum_rule}
 \end{equation}

Since each 2D unit cell of the BLWC contains two electrons, the 
dynamical matrix 
is a $4\times4$ matrix which we write in block form as 
\begin{equation}
{\bf D} = \left(\begin{array}{cc} 
{\bf D}^{AA} & {\bf D}^{AB} \\
{\bf D}^{BA} & {\bf D}^{BB}
\end{array}\right),
\end{equation}
where ${\bf D}^{AA}$, ${\bf D}^{AB}$, ${\bf D}^{BA}$, and 
${\bf D}^{BB}$ are 
$2\times2$ matrices.
Applying (\ref{gen_dyn_mat}) to the BLWC and using 
translational invariance, we obtain the matrix elements of 
${\bf D}^{AA}$ and ${\bf D}^{AB}$
\begin{eqnarray} 
\left[{\bf D}^{AA}({\bf q})\right]_{\alpha\beta} & = & 
\frac{1}{m_e}\sum_{{\bf R}}\phi_{\alpha\beta}({\bf R})
e^{-i{\bf q}\cdot{\bf R}} , \\
\left[{\bf D}^{AB}({\bf q})\right]_{\alpha\beta} & = & 
\frac{1}{m_e}\sum_{{\bf R}}\phi_{\alpha\beta}({\bf R}-{\bf c})
e^{-i{\bf q}\cdot\left({\bf R}-{\bf c}\right)} ,
\end{eqnarray}
where $m_e$ is the electron mass, and the force constants are
\begin{equation}
\phi_{\alpha\beta}({\bf R})=
\partial_\alpha\partial_\beta\frac{e^2}{R},
\mbox{\hspace{1truecm}${\bf R}\neq0$} , 
\end{equation}
and
\begin{equation}
\phi_{\alpha\beta}({\bf R}-{\bf c})=
\partial_\alpha\partial_\beta
\frac{e^2}{\left[\left|{\bf R}-{\bf c}\right|^2+d^2\right]^{1/2}} . 
\end{equation}
Using Eq.~(\ref{sum_rule}), we find the force constant for ${\bf R}=0$ 
\begin{equation}
\phi_{\alpha\beta}({\bf R}=0)=-\left[\sum_{{\bf R}\neq0} 
\phi_{\alpha\beta}({\bf R})
+\sum_{{\bf R}}\phi_{\alpha\beta}({\bf R}-{\bf c})\right].
\end{equation}
Furthermore, since the two sub-lattices are equivalent, 
we have ${\bf D}^{AA}={\bf D}^{BB}$ and, using 
Eq.~(\ref{gen_dyn_mat}), ${\bf D}^{AB}=[{\bf D}^{BA}]^\dagger$. 

It turns out to be  convenient to define
\begin{eqnarray}
\left[{\bf S}^{AA}({\bf q})\right]_{\alpha\beta}
&=&-e^2\sum_{{\bf R}\neq0}\partial_\alpha\partial_\beta
\frac{e^{-i{\bf q}\cdot{\bf R}}}{R} , \\
\left[{\bf S}^{AB}({\bf q})\right]_{\alpha\beta}
&=&-e^2\sum_{{\bf R}}\partial_\alpha\partial_\beta
\frac{e^{-i{\bf q}\cdot\left({\bf R}-{\bf c}\right)}}
{\left[\left|{\bf R}-{\bf c}\right|^2+d^2\right]^{1/2}},
\end{eqnarray}
which can be obtained from $T_0$ and $T_I$
\begin{eqnarray}
\left[{\bf S}^{AA}({\bf q})\right]_{\alpha\beta}&=&
-e^2\lim_{r\rightarrow0}
\partial_\alpha\partial_\beta T_0\left({\bf r},{\bf q}\right), \\
\left[{\bf S}_{AB}({\bf q})\right]_{\alpha\beta}&=&-e^2
\partial_\alpha\partial_\beta T_I\left(0,{\bf q}\right). 
\end{eqnarray}

Then the matrix elements of the dynamical matrix can be written
\begin{eqnarray}
{\bf D}^{AA}({\bf q})&=&\frac{1}{m_e}\left[
{\bf S}^{AA}\left(0\right)+{\bf S}^{AB}\left(0\right)
-{\bf S}^{AA}({\bf q})\right], \\
{\bf D}^{AB}({\bf q})&=&\frac{1}{m_e}\left[
-{\bf S}^{AB}({\bf q})\right]. 
\end{eqnarray}
Using the rapidly convergent form for $T_0$ and $T_I$, as given in 
(\ref{T0}) and (\ref{Ti}), allows one to write down the matrix 
elements of ${\bf S}^{AA}$ and ${\bf S}^{AB}$ explicitly
\widetext
\begin{eqnarray} 
\left[{\bf S}^{AA}({\bf q})\right]_{\alpha\beta}&=&\sqrt{n_s}\left\{
-\sum_{\bf G} 
\left({\bf q}+{\bf G}\right)_\alpha \left({\bf q}+{\bf G}\right)_\beta
\Phi\left(\frac{\left|{\bf q}+{\bf G}\right|^2}{4\pi n_s}\right)
\right.\nonumber\\ &&\left.
+\sum_{{\bf R}\neq0}\overline{V}_{\alpha\beta}(\pi n_s 
R^2)e^{-i{\bf q}\cdot{\bf R}}
+\delta_{\alpha\beta}\frac{4}{3}(\pi n_s) \right\} , \\
\left[{\bf S}^{AB}({\bf q})\right]_{\alpha\beta}&=&\sqrt{n_s}\left\{
-\sum_{\bf G} \left({\bf q}+{\bf G}\right)_\alpha 
\left({\bf q}+{\bf G}\right)_\beta
\Psi\left(\frac{\left|{\bf q}+{\bf G}\right|^2}{4\pi n_s},
\pi\eta^2\right)
e^{-i{\bf G}\cdot{\bf c}}
\right.\nonumber\\ &&\left.
+\sum_{\bf R} \overline{V}_{\alpha\beta}(\pi n_s |{\bf R}-{\bf c}|^2)
e^{-i{\bf q}\cdot\left({\bf R}-{\bf c}\right)} 
\right\} ,
\end{eqnarray}
\narrowtext\noindent
where we have defined
\begin{equation}
\overline{V}_{\alpha\beta}(X^2)=\partial_\alpha\partial_\beta 
\Phi(X^2) 
=2\pi n_s\left\{\delta_{\alpha\beta}\Phi^\prime\left(X^2\right)
+2\pi n_s X_\alpha X_\beta\Phi^{\prime\prime}\left(X^2\right)\right\}.
\end{equation}
\narrowtext

In general ${\bf D}$ is a complex hermitian matrix. 
However, since in the BLWC the two sites of each cell are 
occupied by identical particles, it is 
possible to apply a unitary transformation which results in a real 
symmetric matrix.~\cite{WallaceTC} If we denote with ${\bf I}_2$ the 
$2\times2$ identity matrix, the transformation 
\begin{equation}
{\bf U}=\frac{1}{\sqrt{2}}\left(\begin{array}{cc}
{\bf I}_2 & i{\bf I}_2 \\
i{\bf I}_2 & {\bf I}_2
\end{array}\right)
\end{equation}
results in
\begin{equation}
\overline{{\bf D}}={\bf U}{\bf D}{\bf U}^{-1}  
=\left(\begin{array}{cc}
{\bf D}_{AA}+\mbox{Im}\,{\bf D}_{AB} & \mbox{Re}\,{\bf D}_{AB} \\
\mbox{Re}\,{\bf D}_{AB} & {\bf D}_{AA}-\mbox{Im}\,{\bf D}_{AB}
\end{array}\right),
\label{Dreal}
\end{equation}
where $\mbox{Re}\,{\bf D}_{AB}$ and $\mbox{Im}\,{\bf D}_{AB}$ are the 
real and imaginary parts of ${\bf D}_{AB}$. Note that 
$\mbox{Im}\,{\bf D}_{AB}=0$ for a lattice with inversion symmetry. 
This applies to all phases, except for phase V.

Finally, we solved the set of four linear equations 
\begin{equation}
\left(\overline{{\bf D}}({\bf q})-\omega^2_{{\bf q},j}{\bf 
I}_4\right){\bf e}({\bf q},j)=0 , \label{diag}
\end{equation}
where ${\bf I}_4$ is the $4\times4$ identity matrix, 
$\omega_{{\bf q},j}$   is the frequency of the $j$-th phonon mode 
($j=1,\dots,4$) 
with wavevector ${\bf q}$, and ${\bf e}({\bf q},j)$ its eigenvector.
Equation~(\ref{diag}) is 
equivalent to the diagonalization of the $4\times4$ matrix 
$\overline{{\bf D}}$, which  
provides  the four eigenvalues $\omega^2_{{\bf q},j}$ at each 
point ${\bf q}$ in reciprocal space. For a lattice to be stable, 
it is necessary that $\omega^2_{{\bf q},j}>0$.

Figure \ref{fig:phonons} shows the frequencies $\omega_{{\bf q},j}$
(or, when $\omega^2_{{\bf q},j}$ is negative, its imaginary part) for
phases: (a) I, (b) III, and (c) V, and their evolution with $\eta$. 
Frequencies are given in terms of the characteristic frequency 
$\omega_1=e^2 n^{3/2}/m_e$, which depends on the density and not 
on the lattice geometry. Phonon
dispersions are shown along the high symmetry directions in 
reciprocal space. The 
high symmetry points are labeled according to the insets. We 
recall that in a SLWC the transverse acoustical (TA)  
and the longitudinal 
acoustical (LA) modes vanish at the $\Gamma$ point as $q$ and 
$q^{1/2}$, 
respectively.~\cite{Bonsall77} Thus the sound velocity of the
LA mode is infinite.  The latter behaviour is a general
property of a 2D Coulomb plasma~\cite{Bonsall77} and does not
depend on the lattice geometry, nor on $d$, as is clear from a
comparison of the three panels in Fig.~\ref{fig:phonons}. 
The remaining two (optical) modes ensuing from (\ref{diag}) are 
peculiar to the BLWC and correspond to out-of-phase vibrations of 
electrons in opposite layers. 

Starting from the top panel, it is shown in 
Fig.~\ref{fig:phonons} that, 
as $\eta$ is increased, the TA mode of phase I 
softens until, above a critical value of $\eta$, the 
frequency becomes imaginary, indicating a lattice instability. 
For $\eta$ between $0.262$ 
and $0.622$ phase III (square lattice) is energetically favoured, 
according to Fig.~\ref{fig:totene2}. We recall that 
in a SLWC the square lattice 
cannot exists, since it has an imaginary TA branch.~\cite{Bonsall77} 
Figure~\ref{fig:phonons}(b) shows, on the other hand, that in the 
BLWC the square 
lattice is stable for a certain range of $\eta$. For 
$\eta$ below a critical value, however, the TA branch softens
along the ${\Gamma\mbox{M}}$ direction
and eventually becomes imaginary, as is expected 
from the fact that in this limit the  BLWC tends to 
the SLWC; at the opposite limit of large $\eta$, the TA branch 
softens along the ${\Gamma\mbox{X}}$ direction and eventually phase 
III becomes unstable, as the BLWC tends to two separated SLWC's.

Phase V is stable for large $\eta$ [Fig.~\ref{fig:phonons}(c)]. In 
the limit $\eta\rightarrow\infty$ we have
the phonon dispersion curves of two uncoupled SLWCs; therefore, in
Fig.~\ref{fig:phonons}, each curve in the $\eta=\infty$ case is doubly
degenerate and all modes approach zero for $q\rightarrow0$. 
For smaller $\eta$, optical modes with a
finite frequency at the $\Gamma$ point appear; at the same time, the
TA mode becomes softer and, eventually, becomes
imaginary at $\eta\sim0.6$. 

The sound velocity of the TA mode, $v_{TA}=d\omega_{TA}/dq|_{q=0}$,
along the in-plane directions $(1,0)$ and $(1,1)$, and for the 
five phases, is shown in detail in Fig.~\ref{fig:sound}. The 
labels on top of the figure indicate the energetically favoured 
phase in each range of $\eta$, according to 
Figs.~\ref{fig:totene} and \ref{fig:totene2}; phase I is favoured 
only in a very small range around $\eta=0$ and, therefore,  is not 
indicated.
The vanishing of the low-frequency modes in certain directions, 
shown in Fig.~\ref{fig:sound}, 
sets a limit to the range of stability of each phase. Note 
that phases II and III have a soft mode at a 
value of $\eta$ which coincides with the value where the 
transition between the two phases takes place (the vertical 
dotted line). The same happens for phases III and IV. Therefore, 
the range where phase III is energetically favoured coincides with 
the range of stability of this phase. This has profound 
implications in determining the DLWC phase diagram at $T\neq0$, as we 
will show in the next section. Note also that in the range of 
$\eta$ where phase II is
energetically favoured, both phases I and II are stable, 
i.e., they do not have imaginary phonon frequencies.  
Analogously, in the range
where phase IV is energetically favoured, both phases IV and  
V are stable. The T=0 phase
diagram, deduced from Figs.~\ref{fig:totene} and \ref{fig:totene2}, and
the range of stability, deduced from the softening of phonon modes,
are summarized in Fig.~\ref{fig:phase0}. 

Figure ~\ref{fig:optical} shows the evolution of the optical 
frequencies at the 
$\Gamma$ point, $\omega_{opt}$, with $\eta$. For phases III and V 
the two optical frequencies 
are degenerate. It has been noted that the detection of 
the exponential decay of 
the optical modes at large $\eta$ could serve as a fingerprint of 
the solid phase in a bi-layer structure.~\cite{Falko94}
Note also the different behaviour of the optical modes 
between phase I and II, and between IV and V, which may be 
used experimentally to distinguish between the different possible 
phases.

The dependence of the sound velocity and of 
the optical modes at the $\Gamma$ point upon $\eta$ has been fitted 
to simple analytical expressions in the low $\eta$ (below 
$\sim0.2$) and in the large $\eta$ (above $\sim0.7$) range. 
The sound velocity $v_{TA}$ (see Fig.~\ref{fig:sound}) has been 
fitted to
\begin{equation}
v_{TA}\sqrt{n}/\omega_1 = p_0+p_2\eta^2+p_4\eta^4
\label{fit1_small}
\end{equation}
for small $\eta$, and to 
\begin{equation}
v_{TA}\sqrt{n}/\omega_1 = p_0+p_2\eta^{-2}+p_4\eta^{-4}
\label{fit1_large}
\end{equation}
for large $\eta$. The coefficients $p_i$ for 
phases I, II (small $\eta$) and V (large $\eta$) 
are reported in Table~\ref{fit1}.
The frequencies $\omega_{opt}$ (see 
Fig.~\ref{fig:optical}) have been fitted to 
\begin{equation}
\omega_{{opt}}/\omega_1 = q_0+q_2\eta^2+q_4\eta^4
\label{fit2_small}
\end{equation}
for small $\eta$, and to
\begin{equation}
\omega_{{opt}}/\omega_1 = q_1 e^{-q_2\eta}
\label{fit2_large}
\end{equation}
for large $\eta$. For $\eta\sim 0.262$, close to the boundary 
between phase II and phase III, 
the optical modes of phase II have a singular behaviour. In this 
range we have fitted $\omega_{{opt}}$ to 
\begin{equation}
\omega_{{opt}}/\omega_1 = q_0+q_1(q_2-\eta)^{q_3}.
\label{fit2_boundary}
\end{equation}
 The coefficients $q_i$ are reported in Table~\ref{fit2}. The
agreement with the full calculations did not improve by adding odd
powers of $\eta$ in Eqs.~(\ref{fit1_small}), (\ref{fit1_large}),
and (\ref{fit2_small}). All the above fitting functions give the
correct values with an accuracy better than $0.6\,\%$ in the 
relevant ranges of $\eta$. 

In Fig.~\ref{fig:DOS} we report the evolution with $\eta$ of the  
phonon density of state (DOS).
At each value of $\eta$, we show the DOS of the 
phase which is energetically favoured at that value.
At $\eta=0$ and $\eta=\infty$ the energetically favoured lattice 
are, respectively, the SLWC lattice with density $n$ (phase I), and
two uncoupled SLWCs with density $n/2$ (phase V); 
therefore, the corresponding DOS curves are
equal up to a factor $2^{3/4}$ in the frequency scale. Note in
Fig.~\ref{fig:DOS} the  peak of optical frequencies which narrows at
$\eta\sim0.5$, corresponding to the range of $\eta$ where the
in-plane component of the average interaction of one particle with
its NN's in the same layer and in the opposite layer
are similar. Also note the low-frequency peak which moves to very
low frequencies around $\eta\sim 0.3$ and $\eta\sim 0.7$. This
behaviour is reminiscent of the softening of the TA mode of
the square lattice (phase III) discussed above. The resulting high 
density of low-frequency
modes suggests that very large fluctuations of particles around their
equilibrium lattice sites are possible; correspondingly, a low
melting temperature is expected in proximity of these points, as 
will be discussed in Sec.~\ref{sec:finiteT}.

\section{phase diagram and melting}
\label{sec:finiteT}

In this section we will be concerned with the non-zero temperature
properties of the BLWC. First, we will use the calculated phonon
excitation frequencies to estimate the melting temperature $T_M$
via the Lindemann criterion. In principle, only order of magnitude
estimates of $T_M$ are expected from an harmonic theory, since
anharmonic terms of the potential become important when crystal
vibrations are so large that the lattice is near to dissolve. In
the case of the SLWC, apart from simulations, analytical methods
have been successfully used to calculate $T_M$ by including
anharmonic effects.~\cite{Fisher82,Saitoh89} These methods assume
that melting proceeds through the dislocation mediated mechanism
proposed by Kosterlitz and Thouless,~\cite{Kosterlitz73} Halperin
and Nelson,~\cite{Halperin78} and Young~\cite{Young79} (KTHNY
theory). The ingredient of these calculations is the sound
velocity of the TA mode of the lattice, which is assumed isotropic
in the KTHNY theory and which is indeed the case in the simple
hexagonal lattice of the SLWC, but not in the BLWC, where the TA
mode, in general, is anisotropic, as is clear from
Fig.~\ref{fig:sound}. Therefore, in this work we will rely on the
simple Lindemann criterion. We shall see that, taking the
Lindemann parameter $\delta$, defined in Eq.~(\ref{lindemann})
below, from existing simulations, effectively includes anharmonic
effects into the theory to some extent. 

The Lindemann criterion states that, in a lattice of density 
$n$, melting occurs when \begin{equation}
\frac{\langle u^2\rangle}{r_0^2}=\delta^2 ,
\label{lindemann}
\end{equation}
i.e., when the mean square displacement
of a lattice site around its equilibrium position $\langle
u^2\rangle$  exceeds a 
certain fraction of the mean inter-particle distance 
$r_0=1/\sqrt{\pi n}$. 
The brackets $\langle\dots\rangle$ 
represent the thermodynamic average; in our case the latter will 
be calculated within the harmonic theory. The parameter 
$\delta$ is an input to the criterion,  to be obtained from 
simulations or from some analytic theory. Equation 
(\ref{lindemann}) has been verified in simulations of several 3D 
systems.~\cite{Robbins88} It is known, however, that $\langle 
u^2\rangle$ is logarithmically divergent in 2D. On the other 
hand, the {\em relative} mean square displacement 
$\langle\left|{\bf u}({\bf R})-{\bf u}({\bf R}+{\bf 
a})\right|^2\rangle$, where 
${\bf u}({\bf R})$ and ${\bf u}({\bf R}+{\bf a})$ are the 
displacement vectors at 
lattice site ${\bf R}$ and at the NN site ${\bf R}+{\bf a}$, 
where ${\bf a}$ is the vector joining two NN's, is finite.
Correspondingly, a {\em  modified} Lindemann criterion can be defined
\begin{equation}
\frac{\langle\left|{\bf u}({\bf R})-{\bf u}({\bf R}+{\bf 
a})\right|^2\rangle}{r_0^2} =\delta^2_{m}.
\label{mod_lindemann}
\end{equation}
 The value of $\delta^2_{m}$ at melting has been calculated in
simulations of melting in a SLWC and turned out to be 
$\simeq0.1$.~\cite{Bedanov85} In
principle, $\delta_{m}$ may depend on the lattice geometry and the
nature of the interaction; however, the Lindemann parameter has
been found to be quite independent from the form of the interaction 
both in 2D~\cite{Bedanov85} and in 3D~\cite{Robbins88} systems; 
therefore, we take $\delta^2_{m}=0.1$, and independent from
$\eta$ and from the lattice geometry. Small variations of 
$\delta_m$ would not change our results qualitatively.

The correlation function 
$\langle\left|{\bf u}({\bf R})-{\bf u}({\bf R}+{\bf 
a})\right|^2\rangle$
is calculated within the harmonic theory.~\cite{MaradudinSSP}
Each lattice site in the BLWC has two types of NN's, in general at a
different distance, and the number and distance of the NN's 
changes in a continuous way with $\eta$. Accordingly, we 
calculate separately two (in general different) correlation 
functions
\widetext
\begin{eqnarray}
L_1 & = & \frac{1}{M_1}\sum_{\alpha=x,y}\sum_{m=1\dots M_1}
\langle\left|u^A_\alpha(0)-u^A_\alpha(m)\right|^2\rangle 
\nonumber\\
& = & \frac{4k_BT}{Nm_eM_1}\sum_{{\bf q} j} 
\frac{[e^A_x({\bf q},j)]^2+[e^A_y({\bf q},j)]^2}{\omega^2_{{\bf q},j}}
\sum_{m=1\dots M_1}\sin^2\frac{{\bf q}\cdot{\bf R}_m}{2} , \\
L_2 & = & \frac{1}{M_2}\sum_{\alpha=x,y}\sum_{m=1\dots M_2}
\langle\left|u^A_\alpha(0)-u^B_\alpha(m)\right|^2\rangle \nonumber\\
&=& \frac{k_BT}{Nm_eM_2}\sum_{{\bf q} j} 
\frac{1}{\omega^{2}({\bf q},j)}\sum_{m=1\dots M_1}
\left\{1-2\left[e^A_x({\bf q},j)e^B_x({\bf q},j)+e^A_y({\bf 
q},j)e^B_y({\bf q}, j)\right] \cos{\bf q}\cdot{\bf R}_m\right\} ,
\end{eqnarray}
\narrowtext\noindent
where $k_B$ is the Boltzmann constant, $u^{A(B)}_\alpha$ is the 
$\alpha$-th component of the displacement vector in layer A(B) 
calculated at the origin (0) or at the position of the $m$-th NN.  
$e^{A(B)}_\alpha({\bf q},j)$ is the $\alpha$-th component  of the  
eigenvector of the $j$-th mode, at point ${\bf q}$,
relative to the sub-lattice in layer A(B), ${\bf R}_m$ is the 
relative lattice vector connecting one site to its $m$-th NN in 
the same 
($L_1$) or in the opposite ($L_2$) layer, and the sums over $m$ are 
extended to the $M_1$ ($M_2$) NN's in the same (opposite) sub-lattice.

Now we consider two limiting cases. For $\eta=0$, 
$\langle\left|{\bf u}({\bf R})-{\bf u}({\bf R}+{\bf 
a})\right|^2\rangle=L_1+L_2$, 
since all NN's are equivalent. At the opposite limit, 
$\eta\rightarrow\infty$, 
$\langle\left|{\bf u}({\bf R})-{\bf u}({\bf R}+{\bf 
a})\right|^2\rangle=L_1$, since 
the dynamics in one layer is not influenced by the sub-lattice on 
the opposite layers. Therefore, we write in general
\begin{equation}
\langle\left|{\bf u}({\bf R})-{\bf u}({\bf R}+{\bf 
a})\right|^2\rangle=L_1+f(\eta)L_2 , \label{D}
\end{equation}
where the function $f(\eta)$ satisfies
\begin{equation}
f(0)=1,\,\,\,\,\,f(\infty)=0.
\label{conditions}
\end{equation}
As $f(\eta)$ represents the influence of the oscillation in one 
layer on the oscillations in the opposite layer, we take 
$f(\eta)$ proportional to the in-plane component of the Coulombic
force between two NN sites sitting in opposite layers. 
This is 
\begin{equation}
F_\parallel(d)=-\frac{e^2 c}{\left(c^2+d^2\right)^{3/2}}
\end{equation}
where $c=|{\bf c}|$. Taking $f(\eta)$ proportional to 
$F_\parallel(d)$, and imposing the conditions 
(\ref{conditions}),  we have 
\begin{equation}
f(\eta) = \frac{1}{\left(1+\alpha_p\eta^2\right)^{3/2}} ,
\end{equation}
 where $\alpha_p=\left(nc^2\right)^{-1}$ is a dimensionless
geometric factor which can be calculated from 
Table~\ref{geometries}. 

Inserting Eq.~(\ref{D}) in (\ref{mod_lindemann}) 
and using $r^2_0=1/\pi n_s$, we have calculated the melting 
temperature $T_M$, which is reported in Fig.~\ref{fig:melting} 
for the five phases. For the ``soft'' phases II and IV, $T_M$ was
calculated taking the $T=0$ value of the aspect ratio and $\theta$, 
respectively. This will be justified later.

In the studies of melting of the SLWC, the melting temperature is
usually given in terms of the dimensionless parameter
$\Gamma_M=e^2\sqrt{\pi n}/KT_M$ (the inverse of the vertical units
in Fig.~\ref{fig:melting}), the ratio between the average
Coulombic potential energy and the average kinetic energy. 
Experiments~\cite{Grimes79} give $\Gamma_M\simeq131$ and
simulations~\cite{Morf79} give $\Gamma_M\simeq128$.  Using the
harmonic value of the sound velocity at $T=0$, the KTHNY theory
gives $\Gamma_M\simeq79$. Our calculation, which is performed within
the harmonic approximation, but uses $\delta_{m}$ taken from
simulations which, of course, include anharmonic effects, gives
$k_BT_M/e^2\sqrt{\pi n}=0.00925$ at $\eta=0$, corresponding to
$\Gamma_M=108$. Therefore, our calculation, although overestimates
$T_M$, partially includes anharmonic effects. In a full
anharmonic theory, $L_1$ and $L_2$, which in the harmonic
approximation scale linearly with $T$, would increase more 
rapidly, especially close to the melting transition.

Fig.~\ref{fig:melting} shows that the melting temperature has an
oscillating behaviour as a function of $\eta$.  This is a
consequence of the vanishing of the TA phonon modes at the phase
boundaries II/III and III/IV, as discussed in
Sec.~\ref{sec:dynamic}.  Therefore, for fixed $T\neq0$ and as
function of $\eta$ we observe that alternating solid and liquid
phases are possible, and the reentrant solid phase has a different
lattice geometry each time. Furthermore, note from the inset of
Fig.~\ref{fig:melting} that, for large values of $\eta$, $T_M$
approaches the melting temperature of a SLWC of density $n/2$ from
below.  In certain experimental realizations of the BLWC it could
be easier to change $\eta$ through a change in the charge density,
keeping $d$ constant. Therefore, in Fig.~\ref{fig:melting2} we
show the calculated melting temperature in units of $k_B T_M
d/e^2\sqrt{\pi}$.  Note that in the classical regime the phase
diagram is determined by $\eta$ and a dimensionless temperature,
either $k_BT_M/e^2\sqrt{\pi n}$ or $k_B T_M d/e^2\sqrt{\pi}$.  In
the quantum regime, instead, the kinetic energy term depends on
the density alone and, therefore, the phase diagram must be drawn
explicitly in the tree-parameter space $(d,n,T)$. 

The presence of different lattice geometries which are stable 
within the same 
range of $\eta$ suggests the possibility that, increasing $T$ at fixed 
 $\eta$, the BLWC undergoes a structural phase transition, and, 
eventually, melts at a temperature appropriate to the high 
temperature phase. For example, it seems possible that for 
$\eta<0.262$ the BLWC evolves from phase II  (with some 
value of the aspect ratio 
which minimizes the static energy at $T=0$)  to phase I (aspect 
ratio $\sqrt{3}$), as $T$ exceeds some 
critical value, and eventually  melts at a $T_M$ appropriate for 
phase I. To 
investigate such possibility we have minimized the free energy with 
respect to the lattice geometry at fixed $\eta$ and $T$. The harmonic 
approximation of the free-energy in the high-temperature limit is 
\begin{equation}
F(\xi)=E(\xi)+k_BT\sum_{{\bf q},j} 
\log \frac{\hbar\omega_{{\bf q},j}(\xi)}{k_BT},
\label{free_energy}
\end{equation}
where $\xi$ is a parameter
which defines a distortion of the lattice.  There are two ranges
of $\eta$ where more than one phase is stable with respect to 
lattice vibrations (see
Figs.~\ref{fig:sound} and \ref{fig:phase0}); in the range 
$0.006<\eta<0.262$ phase II 
is energetically favoured, but also phase I is stable throughout 
this 
range. Therefore, in this range, we minimize F with respect to 
$\xi=a_2/a_1$.
In the range $0.622<\eta<0.732$ phase IV is energetically 
favoured, but also phase V is stable; therefore, in this range 
we take $\xi=\theta$. Integration over reciprocal space in
(\ref{free_energy}) was performed numerically.  We found that in
both ranges the value of $\xi$ which minimizes F is practically
independent of the temperature and, therefore, coincides with the
$T=0$ value. In other words, the phase boundaries between the
different geometries in Fig.~\ref{fig:melting} are represented by
vertical lines. Moreover, this justifies the fact that, in order to 
calculate $T_M$ for the ``soft'' phases II and IV, we have used 
the $T=0$ value for the aspect ratio and $\theta$, respectively.

\section*{Conclusions} 
\label{sec:conclusion}

The phase diagram of a classical BLWC, both at $T=0$ and at
$T\neq0$ was investigated, within an harmonic approach, by use of 
the Lindemann criterion and 
minimization of the harmonic free-energy. Five 
different crystalline geometries are stable in different ranges of
inter-layer distance/charge densities. Moreover, at $T=0$ the five
phases evolve one into the other through both continuous and
discontinuous transitions.  At $T\neq0$, alternating solid and
liquid phases are possible, as one sweeps the inter-layer distance
or the charge density. In particular, regions of liquid phase 
separate phase II from phase III, and phase III from phase IV. This 
has been shown to be a consequence of
lattice instabilities induced by the vanishing of phonon modes at
the phase boundaries. On the other hand, a first order transition 
line separates IV from phase V.

An additional intricacy of the phase diagram in the small $\eta$
range has been pointed out by Vil'k and Monarkha~\cite{Vilk85}. In
this limit the hamiltonian (\ref{potential_energy}) was
mapped into the hamiltonian of a binary mixture of particles
sitting on a triangular lattice and interacting through a dipole
potential.  Therefore there is a possibility that a disordered
phase appears, as the temperature is increased. They find two
phase transitions.  The low tempeature (ordered) phase is
equivalent to our phase I. Above a critical temperature $T_1$ the
lattice can be seen as composed of three inter-penetrating triangular
lattices, two of which are ordered and one is disordered. Above a
second critical temperature $T_2$ the lattice becomes completely
disordered. Of course the order-disorder transition vanishes as
$\eta\rightarrow0$, where the two sublattices become equivalent. 

A bi-layer electron gas can easily be realized in semiconductor
heterostructures.~\cite{Suen92,Eisenstein92} Although our results
have been obtained for a classical system, they can give some
indications on the phase diagram in quantum bi-layer structures,
provided that temperature fluctuations are interpreted as quantum
fluctuations. Very recently, in Ref.~\onlinecite{Esfarjani95} a
re-entrant phase around $\eta\simeq2.6$, analogous to ours in
Fig.~\ref{fig:melting}, was predicted in the $(\eta,r_s)$ phase
diagram, where $r_s$ is the dimensionless inverse electron
density.  Furthermore, our analysis of the phonon excitations and
the analytical fitting that we have developed retain their
validity in the quantum regime. 

In principle, the harmonic
approximation used throughout this work is expected to fail when the
temperature approaches the melting temperature. However, we have
shown that, in the $\eta=0$ case, we obtain $T_M$ in reasonable
agreement with numerical simulations and experiments on the SLWC; 
therefore, we believe that inclusion of the anharmonic effects
would not change our results qualitatively, as far as the 
melting temperature is concerned. We believe also that 
the approximation of a structure independent parameter $\delta_m$
in the Lindemann criterion should not change the nature of our
findings. The harmonic approach could be a more severe
approximation in the calculation of the free-energy and our
investigation, therefore, does not rule out completely the
possibility of temperature-induced structural phase transitions. 

\acknowledgments

We acknowledge financial support from the HCM network  
No.~ERBCHRXCT930374, a NATO Collaborative Research Grant, and the 
Belgian National Science Foundation. Discussions with V. A. 
Schweigert, K. Michel, J. Naudts, I. Kono, M. Saitoh and Yu. P. 
Monarkha are gratefully acknowledged.

\appendix

\section{rapidly convergent form of $T_0$ and $T_I$}
\label{sec:appA}

The slowly convergent sums over lattice sites appearing in the 
definition of $T_0$ 
and $T_I$ [Eqs.~(\ref{def_t0}) and (\ref{def_ti})] cannot 
be used in a numerical calculation. Therefore, they will be 
converted into a rapidly convergent form using a 
generalization of the Ewald method.~\cite{Fisher82} Formally, we 
proceed as follows. First, each 
term in the sum is decomposed in two terms, using the identity
\begin{equation}
\frac{1}{r} =\frac{1}{r}  
\left\{\,\mbox{erf}(\varepsilon r)+\,\mbox{erfc}(\varepsilon 
r)\right\}, 
\label{erf_dec}
\end{equation} 
where
\begin{equation}
\,\mbox{erf}(x)=\frac{2}{\sqrt{\pi}}\int_0^x e^{-t^2}\,dt ,
\end{equation}
is the error function, $\,\mbox{erfc}(x)=1-\,\mbox{erf}(x)$ is the 
complementary error function,  and $\varepsilon$ is an arbitrary 
constant. The reason why we do so,  is that  
$\,\mbox{erfc}(x)$ vanishes exponentially for large values of the 
argument and, consequently, the lattice sum with this function as 
argument is sufficiently rapidly convergent.
Then, the  other lattice sum with argument $\,\mbox{erf}(x)$ is mapped 
onto a 
sum over the reciprocal lattice, using the 2D theta-function 
transformation.~\cite{Ziman}

Using (\ref{erf_dec}) and the definition of $T_0$ 
[Eq.~(\ref{def_t0})], we obtain 
\begin{eqnarray}
T_0&=&e^{-i{\bf q}\cdot{\bf r}} \sum_{\bf R} 
\frac{e^{i{\bf q}\cdot\left({\bf r}-
{\bf R}\right)}}{\left|{\bf r}-{\bf R}\right|}
\,\mbox{erf}\left(\varepsilon\left|{\bf r}-{\bf R}\right|\right) 
\nonumber\\
&& + \sum_{{\bf R}\neq0} \frac{e^{-i{\bf q}\cdot{\bf R}}}
{\left|{\bf r}-{\bf R}\right|}
\,\mbox{erfc}\left(\varepsilon\left|{\bf r}-{\bf R}\right|\right) + 
\frac{\,\mbox{erfc}\left(\varepsilon r\right)}{r}-\frac{1}{r}.
\label{t0_split}
\end{eqnarray}
To convert the first sum on the rhs of (\ref{t0_split}) into a 
rapidly convergent 
form, we substitute $\xi=t/\lambda$ in the 
definition of the error function, which results into
\begin{equation}
\frac{\,\mbox{erf}\left(\varepsilon\lambda\right)}{\lambda}
=\frac{2}{\sqrt{\pi}}\int_0^\varepsilon e^{-\lambda^2\xi^2}\,d\xi .
\label{subs}
\end{equation}
with $\lambda=|{\bf r}-{\bf R}|$.
We plug (\ref{subs}) into the first sum of (\ref{t0_split}) and we 
bring the sum under the integral. Next, we apply the 2D 
theta-function transformation~\cite{Ziman}
\begin{equation}
\sum_{\bf R} e^{-\left|{\bf r}-
{\bf R}\right|\xi^2}e^{-i{\bf q}\cdot{\bf R}}
=\frac{n_s\pi}{\xi^2}\sum_{\bf G}
e^{-\left|{\bf q}+{\bf G}\right|^2/4\xi^2}
e^{-i\left({\bf q}+{\bf G}\right)\cdot{\bf r}},
\label{theta}
\end{equation}
and the substitution $t=\left|{\bf q}+{\bf G}\right|/2\xi$, 
which transforms 
the first term on the rhs of (\ref{t0_split}) into
\begin{equation}
2\pi n_s \sum_{\bf G} e^{-i\left({\bf q}+{\bf G}\right)\cdot{\bf r}}
\frac{\,\mbox{erfc}\left(\left|{\bf q}+{\bf 
G}\right|/2\varepsilon\right)} {\left|{\bf q}+{\bf G}\right|}.
\end{equation}
The final step is to choose a reasonable value for $\varepsilon$, 
such that the lattice sums have a sufficient rapid numerical 
convergency. A convenient choice is $\varepsilon= 
r_0^{-1}=\sqrt{\pi n_s}$. Defining the function
\begin{equation}
\Phi(x)  =  \sqrt{\frac{\pi}{x}} \,\mbox{erfc}\left(\sqrt{x}\right)
\end{equation}
to simplify the final expression, we finally obtain Eq.~(\ref{T0}).

For $T_I$ we proceed in a similar way. Let 
$\lambda^2=\left|{\bf r}-{\bf R}+{\bf c}\right|^2+d^2$; 
using (\ref{erf_dec}), and the definition of $T_I$ 
[Eq.~(\ref{def_ti})], we have \begin{equation}
T_I=e^{-i{\bf q}\cdot{\bf r}}\sum_{\bf R} \frac{
e^{i{\bf q}\cdot\left({\bf r}-{\bf R}+{\bf c}\right)}}
{\lambda}
\left[
\,\mbox{erf}\left(\varepsilon\lambda\right)+
\,\mbox{erfc}\left(\varepsilon\lambda\right)
\right].
\label{ti_split}
\end{equation}
Using the identity (\ref{subs}), the 2D theta-function 
transformation 
(\ref{theta}),  and the substitution $t=\left|{\bf q}+{\bf 
G}\right|/2\xi$, the first term in (\ref{ti_split}) becomes
\begin{equation}
\frac{4}{\sqrt{\pi}} \pi n_s
\sum_{\bf G}\frac{e^{-i{\bf G}\cdot{\bf c}} 
e^{-i\left({\bf q}+{\bf G}\right)\cdot{\bf r}}}
{\left|{\bf q}+{\bf G}\right|} 
 \int^\infty_{\left|{\bf q}+{\bf G}\right|/2\varepsilon} 
e^{-d^2 \left|{\bf q}+{\bf G}\right|^2/4t^2} e^{-t^2}\,dt .
\end{equation}
The integral can be performed analytically; using
\begin{equation}
\int^\infty_x e^{-\left(t^2+\alpha^2/t^2\right)}\,dt=
\frac{\sqrt{\pi}}{4}\left[e^{2\alpha}\,
\mbox{erfc}\left(x+\frac{\alpha}{x}\right)+
e^{-2\alpha}\,\mbox{erfc}\left(x-\frac{\alpha}{x}\right)\right],
\end{equation}
inserting $\varepsilon=\sqrt{\pi n_s}$, and defining the function
\widetext
\begin{equation}
\Psi(x,y)  =  \frac{1}{2}\sqrt{\frac{\pi}{x}}
\left[e^{\sqrt{4xy}}\,\mbox{erfc}\left(\sqrt{x}+\sqrt{y}\right) + 
  e^{-\sqrt{4xy}}\,\mbox{erfc}\left(\sqrt{x}-\sqrt{y}\right)\right] ,
\end{equation}
we finally obtain Eq.~(\ref{Ti}).
\narrowtext

\section{Explicit expressions for $E_0$ and $E_I$}
\label{sec:AppB}

The energy $E_0$, calculated from Eq.~(\ref{E_0}), 
contains the divergent term 
\begin{eqnarray}
e^2\sqrt{n_s}\Phi\left(\frac{G^2}{4\pi 
n_s}\right)_{G=0} &=&e^2 \left[\frac{2\pi n_s}{G} - 
\frac{2\pi n_s}{G} \,\mbox{erf}\left(\frac{G}{2\sqrt{\pi n_s}}\right)
\right]_{G=0} \nonumber\\
&=& e^2 n_s \left.\frac{2\pi}{G}\right|_{G=0}-2e^2\sqrt{n_s},
\label{B1}
\end{eqnarray}
where we have made  use of the limit
\begin{equation}
\lim_{x\rightarrow0}x^{-1} \,\mbox{erf}(x) = 2/\sqrt{\pi}.
\label{limit}
\end{equation}
in the second line of (\ref{B1}).
The divergent term in the last line of (\ref{B1}) is independent 
of the lattice geometry and can be neglected. In fact the 
divergency is exactly balanced by
the interaction energy of the electrons 
with a positive background located in the same layer~\cite{Bonsall77}
\begin{equation}
E_0^b = -e^2 n_s \int\frac{d{\bf r}}{r} = 
-\left.e^2 n_s \frac{2\pi}{q}\right|_{q=0} . 
\end{equation}

Equation (\ref{limit}) can also be used to evaluate the 
contribution to $E_0$ of the last two terms in Eq.~(\ref{T0}) 
\begin{equation} 
\lim_{r\rightarrow0}
\left[\sqrt{n_s}\Phi(\pi n_s r^2)-\frac{1}{r}\right]=-2\sqrt{n_s} .
\label{R0}
\end{equation}
Using (\ref{R0}) and  
the identity ${\bf G}= 2\pi n_s \left(\hat{z}\times{\bf R}\right)$, 
where $\hat{z}$ is a unit vector normal to the layers,
$E_0$ reduces finally to Eq.~(\ref{e0_explicit}),
which is equal to Eq.~(2.15) of Ref.~\onlinecite{Bonsall77}. $E_0/2$ 
gives the static energy per electron of a SLWC of density $n_s$.

The divergent term in $E_I$ is 
\widetext
\begin{eqnarray}
&&e^2\sqrt{n_s}\Psi\left(\frac{G^2}{4\pi n_s},\pi\eta^2\right)_{G=0}=
\frac{e^2\pi n_s}{G}
\left\{\left[e^{G\eta/\sqrt{n_s}}+
e^{-G\eta/\sqrt{n_s}}\right]\right.\nonumber\\
&&~~~~~\left.  - \left[
      e^{G\eta/\sqrt{n_s}} \,\mbox{erf}\left(\frac{G}{2\sqrt{\pi n_s}}
                      +\sqrt{\pi}\eta\right)+
      e^{-G\eta/\sqrt{n_s}}\,\mbox{erf}\left(\frac{G}{2\sqrt{\pi n_s}}
                      -\sqrt{\pi}\eta\right) \right]\right\}_{G=0}.
\label{divEi}
\end{eqnarray}
\narrowtext\noindent
The second term on the rhs takes the limit
\begin{equation}
-2e^2\pi n_s\left[\frac{e^{-\pi\eta^2}}{\pi\sqrt{n_s}}+
\frac{\eta}{\sqrt{n_s}}\,\mbox{erf}\left(\sqrt{\pi}\eta\right)\right] ,
\label{Eilimit}
\end{equation}
for $G\rightarrow0$. The first term on the rhs of (\ref{divEi}) 
can be rewritten
\begin{equation}
e^2\pi n_s \left[\right.\frac{2e^{-G\eta/\sqrt{n_s}}}{G}+
\left.\frac{e^{G\eta/\sqrt{n_s}}
-e^{-G\eta/\sqrt{n_s}}}{G}\right]_{G=0}=
e^2 n_s 
\left.\frac{2\pi}{G} 
e^{-G\eta/\sqrt{n_s}}\right|_{G=0}+2e^2\pi\sqrt{n_s}\eta .
\end{equation}
Again, the divergent term on the rhs is independent of the 
lattice geometry and can be neglected. In fact, this term 
is exactly balanced by
the interaction energy of an electron with a positive background 
charge located at the opposite layer
\begin{equation}
E_I^b = -e^2 n_s \int\frac{d{\bf r}}{\left(r^2+d^2\right)^{1/2}} =
-\left.e^2 n_s\frac{2\pi}{k}e^{-k\eta/\sqrt{n_s}}\right|_{k=0},
\label{Eib}
\end{equation}
which balances the divergency. Therefore, we obtain
\begin{equation}
e^2\sqrt{n_s}\Psi\left(\frac{G^2}{4\pi 
n_s},\pi\eta^2\right)_{G=0} = 
2e^2\sqrt{n_s}\left\{\pi\eta\,\mbox{erfc}
\left(\sqrt{\pi}\eta\right)-
e^{-\pi\eta^2}\right\},
\end{equation}
and, finally, Eq.~(\ref{Ei_explicit}).

The background charge does not need to sit in the same
layer of the mobile electrons.  This would be, in fact, the
situation in the 2D electron gas realized in semiconductor
heterostructures, where the positive ions sit far from the
inversion layer. In this case the electrostatic energy has
an additional contribution $2\pi e^2 n_s
(d^\prime+d^{\prime\prime})$, where $d^\prime$, $d^{\prime\prime}$
are the distances between the compensating layers and the electron
layers. Since this additional contribution does not depend on the
inter-layer distance $d$ nor on the lattice structure, it can be
neglected.

%
%
\begin{figure}
 \caption{Static energy per particle of phases I, III, and V. In
the insets we show the corresponding lattice geometries in which 
dots and crosses identify the two sub-lattices. ${\bf a}_1$
and ${\bf a}_2$ are given in Table~\protect\ref{geometries}.} 
 \label{fig:totene}
\end{figure}

\begin{figure}
  \caption{Detail of Fig.~\protect\ref{fig:totene} showing the
transitions: (a) I$\rightarrow$II$\rightarrow$III, and (b)
III$\rightarrow$IV$\rightarrow$V. Also shown are the lattice
geometries in which full dots and crosses identify the two
sub-lattices. Empty dots and diamonds which indicate the
sub-lattices of phases I (a) and III (b) are also reported for
reference. The insets show how: (a) the aspect ratio $a_2/a_1$,
and (b) the sine of the angle $\theta$ between ${\bf a}_1$ and
${\bf a}_2$ evolve during the transition.}
  \label{fig:totene2}
\end{figure}

\begin{figure}
 \caption{Phonon dispersion curves for phase I (top panel), phase 
III (middle panel), and phase V (bottom panel), and for several 
values of $\eta$, as indicated in the legends. Phonon 
frequencies are shown along high symmetry directions in the 
Brillouin zones of the three phases. In each panel, high symmetry 
points along the abscissa are labeled according to the insets. 
Frequencies are given in terms of the characteristic frequency 
$\omega_1^2=e^2n^{3/2}/m_e$.}
 \label{fig:phonons}
\end{figure}

\begin{figure}
 \caption{Sound velocity of the TA mode  
($\omega_1^2=e^2n^{3/2}/m_e$) along the (1,0) (solid 
lines) and (1,1) (dashed lines) directions for the five phases. 
The sound velocity of phase V is isotropic and the two curves 
coincide.
Vertical dotted lines indicate the phase boundaries, according to 
Figs.~\protect\ref{fig:totene} and \protect\ref{fig:totene2};  
the labels on top of the figure indicate  which phase is  
energetically favoured in each region.}
 \label{fig:sound}
\end{figure}

\begin{figure}
  \caption{T=0 phase boundaries (solid dots) and range of stability 
(crosses) of the five phases along the $\eta$ axis.}
  \label{fig:phase0}
\end{figure}

\begin{figure}
 \caption{Optical frequencies at the $\Gamma$ point for the five
phases. For each phase, $\omega_{opt}$ are reported in the
whole range in which the phase is stable. Vertical
dotted lines indicate the phase boundaries; the
labels on top of the figure indicate which phase is energetically
favoured in each region.}
 \label{fig:optical}
\end{figure}

\begin{figure}
  \caption{Phonon DOS as a function of frequency for 
different values of $\eta$. 
For each value of $\eta$, the DOS corresponding to the
energetically favoured lattice is reported. Dashed lines 
indicate that a ``soft'' phase (either II or IV) is stable 
at that value of $\eta$. $\omega_1^2=e^2n^{3/2}/m_e$.}
  \label{fig:DOS} 
\end{figure}

\begin{figure}
  \caption{Melting temperature $T_M$ for the five phases. 
For the ``soft'' phases II and IV (dashed lines) we used
the value of the continuously changing parameter, either the 
aspect ratio $a_2/a_1$ or the angle between ${\bf a}_1$ and 
${\bf a}_2$, 
respectively, for which the energy is at its minimum at T=0.}
  \label{fig:melting} 
\end{figure}

\begin{figure}
  \caption{Same as Fig.~\protect\ref{fig:melting}, but with 
$T_M$ plotted in units of $(k_B d/e^2 \protect\sqrt{\pi})^{-1}$.
}
  \label{fig:melting2} 
\end{figure}
%
%
\widetext
\begin{table}
\caption{Lattice parameters of the five geometries considered. 
$a$ is the NN distance. For each phase, the primitive vectors 
${\bf a}_1$ and ${\bf a}_2$, 
the inter-lattice displacement ${\bf c}$, the reciprocal lattice 
vectors ${\bf b}_1$ 
and ${\bf b}_2$, and the charge density $n_s$ are indicated. For phase 
II,  $a_2/a_1$ is 
the aspect ratio. 
For phase IV, $\theta$ is the angle between ${\bf a}_1$ and 
${\bf a}_2$.} 
\label{geometries} 
\begin{tabular}{lcccccc}
Phase & ${\bf a}_1/a$ & ${\bf a}_2/a$ & ${\bf c}$ & 
${\bf b}_1/(2\pi/a)$ & ${\bf b}_2/(2\pi/a)$ & $n_s a^2$ \\\hline
I--One-component hexagonal & $(1,0)$ & $(0,\sqrt{3})$ & 
$({\bf a}_1+{\bf a}_2)/2$ & $(1,0)$& 
$(0,1/\sqrt{3})$& $1/\sqrt{3}$ \\
II--Staggered rectangular & $(1,0)$ & $(0,a_2/a_1)$ & 
$({\bf a}_1+{\bf a}_2)/2$ & $(1,0)$& 
$(0,a_1/a_2)$& $a_1/a_2$ \\
III--Staggered square & $(1,0)$ & $(0,1)$ & 
$({\bf a}_1+{\bf a}_2)/2$ & $(1,0)$& 
$(0,1)$& $1$ \\
IV--Staggered rhombic & $(1,0)$ & $(\cos\theta,\sin\theta)$ & 
$({\bf a}_1+{\bf a}_2)/2$ & 
$(1,-\cos\theta/\sin\theta)$& 
$(0,1/\sin\theta)$& 
$1/\sin\theta$ \\ 
V--Staggered hexagonal & $(1,0)$ & $(1/2,\sqrt{3}/2)$ & 
$({\bf a}_1+{\bf a}_2)/3$ & $(1,-1/\sqrt{3})$& 
$(0,2/\sqrt{3})$& $2/\sqrt{3}$ 
\end{tabular}
\end{table}
\narrowtext

\begin{table}
\caption{Fitting parameters for the sound velocity $v_{TA}$ in 
Eqs.\protect(\ref{fit1_small}) and (\ref{fit1_large}).} 
\label{fit1} 
\begin{tabular}{c|@{}cc|@{}cc|@{}c}
 & \multicolumn{4}{c|}{Small $\eta$} 
 & Large $\eta$ \\
 & \multicolumn{4}{c|}{[Eq.~(\ref{fit1_small})]} 
 & ~[Eq.~(\ref{fit1_large})]~ \\\cline{2-6}
 & \multicolumn{2}{c|}{Phase I} & \multicolumn{2}{c|}{Phase II} 
 & Phase V \\
 & (1,0) & (1,1)  & (1,0) & (1,1) & \\\cline{2-6}
$p_0$ & \multicolumn{2}{c|}{0.49504} & \multicolumn{2}{c|}{0.49504} 
    & 0.41628 \\
$p_2$ & ~-3.6871 & 0.748 & ~1.6608 & -3.5072 & 0.01832 \\
$p_4$ & ~-6.1097 & -1.0444 & ~1.5212 & -21.192 & -0.05925 
\end{tabular}
\end{table}

\widetext
\begin{table}
\caption{Fitting parameters for the optical frequencies 
$\omega_{opt}$ in Eqs.\protect(\ref{fit2_small}), 
(\ref{fit2_large}), and (\ref{fit2_boundary}).} 
\label{fit2} 
\begin{tabular}{c|@{}cc|@{}cc|c|@{}cc|@{}c}
 & \multicolumn{4}{c|}{Small $\eta$ } &
 & \multicolumn{2}{c|}{Phase boundary } 
 & Large $\eta$ \\
 & \multicolumn{4}{c|}{[Eq.~(\ref{fit2_small})]} &
 & \multicolumn{2}{c|}{[Eq.~(\ref{fit2_boundary})]} 
 & ~[Eq.~(\ref{fit2_large})]~ \\\cline{2-5}\cline{7-9}
 & \multicolumn{2}{c|}{Phase I} & \multicolumn{2}{c|}{Phase II} &
 & \multicolumn{2}{c|}{Phase II} & Phase V \\
 & ~Low branch & High branch &  ~Low branch & High branch &
 & ~Low branch & High branch &  \\\cline{2-5}\cline{7-9}
$q_0$ & 1.0706 & 3.0379 & 1.0706 & 3.0379 &  
$q_0$ & \multicolumn{2}{c|}{1.9549} & \\
$q_2$ & -3.5818 & -7.6876 & 5.8313 & -8.2276 & 
$q_1$ & -2.0696 & 2.6527 & 5.7169 \\ 
$q_4$ & 6.6545 & 12.742 & 38.141 & -44.859 &
$q_2$ & 0.26252 & 0.26252  & 3.2168 \\
 & & & & & $q_3$ & 0.45103 & 0.48989 & \\
\end{tabular}
\end{table}
\narrowtext


\begin{references} 

\bibitem[\ast]{byline1} Present address: Department of Physics,
University of Modena, Via Campi 213/A, 41100 Modena, Italy.
E-mail address: goldoni@unimo.it.

\bibitem[\dag]{byline2} E-mail address: peeters@uia.ua.ac.be. 

\bibitem{Isihara89} See, for example, A. Isihara, Solid State 
Phys.~{\bf 42}, 271 (1989).

\bibitem{Grimes79} C. C. Grimes and G. 
Adams, Phys.~Rev.~Lett.~{\bf 42}, 795 (1979).

\bibitem{Murray87}  C. A. Murray and D. M. Winkle, 
Phys.~Rev.~Lett.~{\bf  58}, 1200 (1987); F. M. Peeters and X. 
Wu, Phys.~Rev.~B {\bf 35}, 3109 (1987). 
 
\bibitem{Clark91} For a recent review, see, for example, R. G. 
Clark, Physica Scripta {\bf T39}, 45 (1991).

\bibitem{Strandburg88} See, for example, K. J. Strandburg, 
Rev.~Mod.~Phys.~{\bf 60}, 161 (1988).

\bibitem{Kosterlitz73}  M. Kosterlitz and D. Thouless, 
J.~Phys.~C {\bf 6}, 1181 (1973).

\bibitem{Halperin78} B. I. Halperin and D. R. Nelson, 
Phys.~Rev.~Lett.~{\bf 41}, 121 (1978); {\em ibid.} E {\bf 41}, 519 
(1978).

\bibitem{Young79} A. P. Young, Phys.~Rev.~B {\bf 19}, 1855 (1979).

\bibitem{Vilk84} Y. M. Vil'k and Y. P. Monarkha, Sov.~J. Low 
Temp.~Phys.~{\bf 10}, 465 (1984) [Fiz.~Nizk.~Temp. {\bf 10}, 886 
(1984)].

\bibitem{Vilk85} Y. M. Vil'k and Y. P. Monarkha, Sov.~J. Low 
Temp.~Phys.~{\bf 11}, 535 (1985) [Fiz.~Nizk.~Temp.~{\bf 11}, 971 
(1985)].

\bibitem{Falko94} V. I. Falko, Phys.~Rev.~B {\bf 49}, 7774 (1994).

\bibitem{Esfarjani95} K. Esfarjani and Y. Kawazoe, J. Phys.: 
Condens.~Matter {\bf 7}, 7217 (1995).

\bibitem{Goldoni95} G. Goldoni, V. A. Schweigert, and F. M. 
Peeters, to appear in Surf.~Sci. (Proceedings of 
the EP2DS-XI conference, Nottingham, August 1995).

\bibitem{NarasimhanPre}  S. Narasimhan and T.-L. Ho 
(unpublished).

\bibitem{ChanPre}  A. Chan and A. H. Mac Donald (unpublished).

\bibitem{Suen92} Y. W. Suen, L. W. Engel, M. B. Santos, M. 
Shayegan, and D. C. Tsui, Phys.~Rev.~Lett.~{\bf 68}, 1379 (1992).

\bibitem{Chen93} X. M. Chen and J. J. Quinn, Phys.~Rev.~B {\bf 
47}, 3999 (1993).

\bibitem{Bedanov85}  V. M. Bedanov, G. V. Gadiyak, Y. E. Lozovik, 
Phys.~Lett.~{\bf 109A}, 289 (1985).

\bibitem{Bonsall77}  L. Bonsall and A. A. Maradudin, Phys.~Rev.~B 
{\bf 15}, 1959 (1977).

\bibitem{Fisher82} D. S. Fisher, Phys.~Rev.~B {\bf 26}, 5009 (1982).

\bibitem{MaradudinSSP} See, for example, A. A. Maradudin, E. W. 
Montroll, G. H.  Weiss, and I. P. Ipatova, {\em Theory of lattice 
dynamics in the harmonic approximation}, in Solid State Physics,
Advances in Research and Applications, Supplement 3, Eds.  H. 
Ehrenreich, F.
Seitz, and D. Turnbull (Academic Press, New York, 1971). 

\bibitem{WallaceTC} See, for example, D. C. Wallace, {\em 
Thermodynamics of crystals} (John Wiley \& Sons, New York, 1972).

\bibitem{Saitoh89} M. Saitoh, Phys.~Rev.~B {\bf 40}, 810 (1989).

\bibitem{Robbins88} M. O. Robbins, K. Kremer, and G. S. Grest, 
J. Chem.~Phys.~{\bf 88}, 3286 (1988).

\bibitem{Morf79} R. H. Morf, Phys.~Rev.~Lett.~{\bf 43}, 931 (1979).

\bibitem{Eisenstein92} J. P. Eisenstein, G. S. Boebinger, L. N. 
Pfeiffer, K. W. West, and S. He, Phys.~Rev.~Lett.~{\bf 68}, 1383 
(1992).

\bibitem{Ziman} J. M. Ziman, {\em Principles of the theory of solids} 
(Cambridge University Press, U.K., 1972).

\end{references}
\end{document}